\documentclass[12pt]{article}

\usepackage{latexsym,epsfig,graphics,epic,eepic,paralist}
\usepackage{epic}
\usepackage{eepic}

\newtheorem{corollary}{Corollary}[section]
\newtheorem{observation}{Observation}

\newtheorem{proof}{Proof}

\def\BibTeX{{\rm B\kern-.05em{\sc i\kern-.025em b}\kern-.08em
    T\kern-.1667em\lower.7ex\hbox{E}\kern-.125emX}}

\newcommand{\dahntab}[1]{
  \newbox\mybok%
  \setbox\mybok=\hbox{\vbox{
      \begin{tabbing}
        #1
      \end{tabbing}%
    }}

  \newdimen\bokwidth%
  \bokwidth=\wd\mybok%
  \newdimen\myl%
  \myl=\textwidth%
  \divide\myl by 2%
  \divide\bokwidth by -2%
  \advance\myl by\bokwidth%
  \vrule width\myl height 0pt depth 0pt%
  \usebox\mybok%
}
\newcommand{\restored}{{\sc Restored}}

\def\implies{\Rightarrow}

\def\qed{\hfill$\Box$\newline\vspace{5mm}}
\newtheorem{lemma}{Proposition}
\newtheorem{definition}{Definition}
\newtheorem{example}{Example}

\newtheorem{theorem}{Theorem}
\def\beginproof{\noindent{\bf Proof.}\quad}
\def\endproof{\qed}

\begin{document}

\author{Gabriel Istrate \thanks{This work has been supported by PN-II ``Parteneriate'' Grant 10028/14.09.2007
from Romanian CNCSIS.}\\
\small eAustria Institute, V. P\^{a}rvan 4, cam 045B\\[-0.8ex]
\small Timi\c{s}oara, RO 300223, Romania\\[-0.8ex]
\small \texttt{gabrielistrate@acm.org}}

\title{Identifying almost sorted permutations from TCP buffer dynamics}
\date{
(Computer system organization; Network Protocols)}

\maketitle

\begin{abstract}

Associate to each sequence $A$ of integers (intending to represent
packet IDs) a sequence of positive integers of the same length
${\mathcal M}(A)$. The $i$'th entry of ${\mathcal M}(A)$ is the size
(at time $i$) of the smallest buffer needed to hold out-of-order
packets, where space is accounted for unreceived packets as well.
Call two sequences $A$, $B$ {\em equivalent} (written $A\equiv_{FB}
B$) if ${\mathcal M}(A)={\mathcal M}(B)$.

We prove the following result: any two permutations $A,B$ of the
same length with $SUS(A)$, $SUS(B)\leq 3$ (where SUS is the {\em
shuffled-up-sequences} reordering measure
\cite{estivill-sorting-survey}), and such that $A\equiv_{FB} B$ are
identical.

The result (which is no longer valid if we replace the upper bound 3
by 4) was motivated by \restored , a receiver-oriented model of
network traffic we introduced in \cite{restored}.
\end{abstract}



\section{Introduction}

The TCP protocol \cite{stevens-tcp} is the fundamental protocol for
computer communications. TCP breaks the information into {\em
packets}, and attempts to maintain a ordered packet sequence to be
passed to the application layer. It accomplishes this by {\em
buffering} packets that arrive out-of-order.

Recent work in the area of network traffic modeling has brought to
attention the significant impact of packet reordering on the
dynamics of this protocol \cite{bellardo02,pathological,laor}. This
has stimulated research (mainly applied, rather than mathematical)
on measuring and modeling reordering
\cite{reordering-ifip05,ietf-draft-reordering}, and on quantifying
the impact of packet reordering on application performance.

In this paper we study a combinatorial problem motivated by modeling
 packet reordering in large TCP traces: suppose that we map a
sequence $A$ of packet IDs into the sequence of integers ${\mathcal
M}(A)$ representing the different sizes of the buffer space
necessary to store the out-of-order packets; we assume that space in
the buffer is reserved (and accounted) for unreceived out-of-order
packets as well. What kind of additional information on the sequence
$A$ is needed to uniquely identify $A$ given ${\mathcal M}(A)$ ?

The problem arose in the context of \restored\ \cite{restored}, a
method for receiver-oriented modeling and compression of large TCP
traces. Previously we showed experimentally \cite{restored} that
\restored\ is able to regenerate sequences similar to the original
sequences with respect to several reordering metrics. One of these
metrics was the {\em reorder density} (RD) from
\cite{reordering-ifip05,ietf-reorder-density,rbd-ccs07}. For RD the
experimental result is somewhat counterintuitive since
\begin{enumerate}
\item \restored\ generates sequence that are (locally)
similar (with respect to mapping ${\mathcal M}$ to the original
sequence.
\item $RD$ can
 take different values on sequences that map to the same sequence
 via ${\mathcal M}$.
\end{enumerate}
Because of this latter property, the fact that the reconstructed
sequences have similar properties with respect to the original
sequence does not follow from the theoretical guarantee 1).

The result in this paper, together with the experimental observation
that over $99\%$ of the traces we previously considered for
benchmarking RESTORED obey the constraint present in our result,
explains why the theoretical inconsistence of RD is not observed in
the ``real-world'' data we employed to benchmark RESTORED.

\section{Preliminaries}

We first give a brief primer on the relevant aspect of the TCP
protocol, \restored\, and the concepts used in the sequel.

The TCP protocol \cite{stevens-tcp} attempts to maintain an ordered
stream of data bytes, identified by an integer called {\em byte ID},
that is effectively communicated through the network by breaking it
down into {\em packets}. The ordering is maintained by {\em
buffering} out-of-order packets. The dynamics of the buffer can be
described in part using several parameters.

\begin{enumerate}
 \item The first parameter is {\em NextByteExpected}, and is the smallest
 index of a data
 byte that has still not been received by the receiver.
\item A second, related, parameter is {\em LastByteRead}, the index of the last
byte processed by the receiver-side application that communicates
through the network via the TCP protocol. Throughout this paper we
will make the simplifying assumption that data is read by the
application as soon as it is ready. In other words NextByteExpected
= LastByteRead+1.
\item Another parameter is {\em LastByteRcvd}, the index of the last
byte that has arrived at the receiver, awaiting processing.
\item {\em RcvWindow}, the size of the {\em receiver window}, is a
receiver-maintained parameter that is meant to provide the sender an
estimate of the available buffer space at the receiver.
\item Finally, {\em RcvBuffer} is a implementation-dependent system constant,
the size of the
receiving buffer.
\end{enumerate}

The functioning of the TCP protocol ensures that these four
parameters are related through the relation (\cite{kurose-ross}
section 3.5):
\begin{equation}\label{maineq}
 \mbox{RcvWindow} = \mbox{RcvBuffer} -
 [\mbox{LastByteRcvd}-\mbox{LaxtByteRead}].
\end{equation}

The term in parantheses on the right-hand side is the actual size of
the TCP receiver-buffer. The measurement takes into account space
reserved (but not necessarily used) for all packets from the first
expected to the last arrived. This is, of course, proportional to
the buffer size measured in packets rather than bytes if it is the
case that all packets have the same size.

 TCP is {\em receiver-driven}: that is, the receiver attempts to
maintain control on the sender flow stream by directing the sender
speed, and {\em acknowledging} the received packets. An
acknowledgment (shortly, ACK) generally consists of the ID of the
{\em first packet that has not yet been received}. Acknowledgement
mechanisms vary from implementation to implementation, and can
entail {\em delayed} or {\em selective} acknowledgments, urgent
retransmission requests, etc. From our standpoint, what is important
that we can associate a sequence of integer ACKs to every sequence
of packet IDs, the sequence of ACKs that would be sent if the
receiver would immediately acknowledge every packet received.

\begin{example}
Consider the following hypothetical sequence of packet IDs:
$A=(4\;\;\;3\;\;\;  2\;\;\;  1 )$. Then the sequence of ACKS is
$ACK(A)=(1\;\;\; 1\;\;\; 1\;\;\;  5)$.
\end{example}

\restored\  \cite{restored} is  Markovian model of large TCP traces
that incorporates information on the dynamics of packet reordering.
It can be used to provide estimates of various measures of quality
of service without making these measurements online, or storing the
entire sequence. Rather, it first ``compresses'' the trace into a
small ``sketch'' that allows regeneration of a TCP trace with
(hopefully) similar characteristics. If needed, we can then perform
a large number of measurements on the regenerated trace.

For the purposes of the present paper, a {\em connection} is simply
a sequence of integers (packet IDs). Suppose that the receiver
observes the following (hypothetical) packet stream

\[
1 \;\;\; 2\;\;\; 3\; \; \; 6 \;\;\; 5 \;\;\; 7 \;\;\; 4\; \; \; 8
\;\;\; 9 \;\;\; 10\; \; \; 12 \;\;\; 13 \;\;\; 14 \;\;\; 11.
\]

In this example packets with IDs $4,5,6,7,12,13,14$ and $11$ arrive
out of order. One can, consequently, classify the received packets
into two categories: those that can be immediately passed to the
application layer, and those that have to be temporarily stored
before delivery. In the example, packets $5$, $6$, and $7$ are
temporarily buffered, and the buffer is only flushed when packet $4$
is received. Similarly, packets $12$, $13$, and $14$ are temporarily
buffered, and the buffer is flushed when packet $11$ arrives. We
will call a packet that marks the end of a sequence of consecutively
buffered packets a {\em pivot packet}. Packets that are immediately
delivered to the application layer are also trivially pivots. In our
example this is the case for packets $1$, $2$, $3$, $4$, $8$, $9$,
$10$ and $11$.

The distinction we introduced effectively defines a coarsened
representation of the stream of packet IDs using two states: An {\em
ordered state} $\mathcal{O}\,$, in which packets arrive when they
were supposed to, and an {\em unordered state} $\mathcal{U}\,$ in
which there is reordering and buffering. Each occurrence of State
$\mathcal{O}$ is followed by one or more occurrences of State
$\mathcal{U}$.

The dynamics of packet IDs in the ordered state is trivial by
definition: in order, starting with the first expected packets. In
\cite{restored} we dealt with the dynamics of packet IDs in the
unordered state, and defined a many-to-one mapping ${\mathcal M}$,
sending sequences of IDs into ``sketches.''

\begin{definition}
Let $A=\{A_1,A_2,\ldots,A_n\}$ be a sequence of packet IDs. We
define the ${\mathcal M}$ as an operator that after receiving a
packet $A_i$ at time index $i$, outputs the difference between the
highest ID ($H_i$) seen so far and the highest ID ($L_i$) that could
be uploaded.
\begin{equation}
{\mathcal M}(A_i) = H_i - L_i. \label{eq1}
\end{equation}
In other words, ${\mathcal M}$ is the size of the smallest buffer
large enough to store all packets that arrive out-of-order, where
the definition of size accounts for reserving space for unreceived
packets with intermediate IDs as well. The \emph{buffer sequence}
${\mathcal M}(P)$ associated with a sequence $P$ of packet IDs is
simply a time-series of ${\mathcal M}$ values.

Two sequences of packet IDs $P$ and $Q$ are {\em full buffer (FB)
equivalent} (written $P\equiv_{\footnotesize\textup{FB}} Q$) if
${\mathcal M}(P)={\mathcal M}(Q)$.
\end{definition}

\begin{example}
Let $A=(4\;\;\;3\;\;\;  2\;\;\;  1 )$. Then ${\mathcal
M}(A)=(4\;\;\; 4\;\;\; 4\;\;\;  0)$.
\end{example}

The mapping ${\mathcal M}$ is many-to-one, but an inverse can be
computed in polynomial time \cite{restored-dam}. This was used in
the regeneration algorithm, where  in the unordered state we first
sample a sketch $S$ from the distribution of such sketches and then
reconstruct a sequence of IDs that maps (via ${\mathcal M}$)  to
$S$.

Mapping ${\mathcal M}$ provides a formal way to guarantee that the
reconstructed sequences are locally ``similar'' to the original one.
The formal notion of similarity has implication for the dynamics of
the TCP protocol:

\begin{definition}\label{def}
Two packet sequences $A,B$ are {\em behaviorally equivalent} if they
yield the same sequence of ACKs.
\end{definition}

Suppose now that a TCP implementation uses simple ACKs (as opposed
to SACK), and acknowledges every single packet then {\em two
sequences that map (via ${\mathcal M}$) to the same sequence are
behaviorally equivalent} \cite{restored-shiva}. As the dynamics of
the congestion window is receiver-driven, assuming identical network
conditions for the two ACK sequences, the two traces can be regarded
as ``equivalent,'' from a receiver-oriented standpoint.

 We will also need a standard measure of
disorder \cite{estivill-sorting-survey}. This measure is denoted by
shuffled up-sequences (SUS) and is defined as follows:

\begin{definition}
Given sequence of integers $A$ denote by $SUS(A)$ {\em the minimum
number of ascending subsequences into which we can partition $A$.}
\end{definition}

For example, a sequence $A = \langle
6,5,8,7,10,9,12,11,4,3,2\rangle$ has
\begin{equation}
\textup{SUS}(A) = \|\{\langle 6,8,10,12\rangle,\langle
5,7,9,11\rangle,\langle 4\rangle,\langle 3\rangle,\langle
2\rangle\}\| = 5,
\end{equation}
where $\|S\|$ denotes the cardinality of a set $S$.

\section{Main result}

In this section we will prove our main result:

\begin{theorem}\label{main}
Let $A,B$ be permutations of length $n$ with $SUS(A),SUS(B)\leq 3$ such that
$A\equiv_{FB}B$. Then $A=B$.
 \end{theorem}

\begin{observation}
 The theorem is no longer true if we replace the condition with
 $SUS(A)$, $SUS(B)\leq 4$. This is witnessed by
 sequences $(4\;\;\; 3\;\;\; 2\;\;\; 1)$ and $(4\;\;\; 2\;\;\; 3\;\;\; 1)$.
 Indeed $A\equiv_{FB}B$, since they both map to sequence
 $(4\;\;\; 4\;\;\; 4\;\;\; 0)$. In fact $SUS(A)=4$, $SUS(B)=3$.
\end{observation}

\beginproof
We consider the greedy algorithm for computing SUS 
displayed in Figure~\ref{alg}. The algorithm has been implicitly
proved correct in \cite{levcopoulos-petersson-sus}; the reason is
parameter SUS was shown to be equal to another presortedness measure
denoted by LDS, and defined as follows:

\begin{definition}
Let $A=(a_{1},a_{2},\ldots, a_{n})$ be a sequence of nonnegative
integers. $LDS(A)$ is defined as the longest length of a decreasing
subsequence $a_{i_{1}}>a_{i_{2}}>\ldots a_{i_{j}}$ ($1\leq
i_{1}<i_{2}<\ldots < i_{j}\leq n$) of $A$.
\end{definition}

With this definition it is easy to see that Algorithm~\ref{alg}
computes parameter LDS (to make the paper self-contained we reprove
this result below).

\begin{figure}
\fbox{ \dahntab{
\ \ \ \ \=\ \ \ \ \=\ \ \ \ \=\ \ \ \ \= \ \ \ \ \=\ \ \ \ \= \\
{\bf Algorithm SUSGreedy(W)} \\
\\
{\bf INPUT}: a list $W=(p_{1},p_{2},\ldots,p_{n})$ of non-negative integers. \\
\\
let $i=1$;\\
let $j=1$;\\
let $L_{1}$ be the empty list;\\
while ($i\leq n$)\{\\
\> add $p_{i}$ to the first list $L_{t}$, $1\leq t\leq j$\\
\> where it can be added while maintaining it sorted;\\
\> if this is not possible \\
\> \> \{ \\
\> \> \> j++; \\
\> \> \> create new list $L_{j}=\{p_{i}\}$; \\
\> \> \}\\
\> i++; \\
\}\\
let $u$ be the number of lists created by the algorithm;\\
\\
OUTPUT $u=LDS(W)=SUS(W)$. \\

}
}
 \caption{Greedy Algorithm for computing SUS}\label{alg}
\end{figure}

We now give a simple algorithm, displayed in Figure~\ref{algrec},
that, given a sequence $W$ of positive integers constructs (if
possible) a permutation $A$ of size $n$ with $SUS(A)\leq 3$ such
that ${\mathcal M}(A)=W$. The proof that the algorithm is correct
will imply the uniqueness of sequence $A$.

\begin{figure}
\fbox{ \dahntab{
\ \ \ \ \=\ \ \ \ \=\ \ \ \ \=\ \ \ \ \= \ \ \ \ \=\ \ \ \ \= \\
{\bf Algorithm RECONSTRUCT} \\
\\
{\bf INPUT}: a list $W=(w_{1},w_{2},\ldots,w_{n})$ of positive integers. \\
\\
let PACKET and ACK be integer vectors of size $n;$\\
with all fields initially equal to $-1$; \\
let $LARGEST=0$;\\
conventionally define $ACK[0]=0$;\\
for ($i=1$ to $n$)\{\\
\>  if ($w_{i}<w_{i-1}$)\{\\
\> \> \> $PACKET[i]=ACK[i-1]$; \\
\> \> \> ACK[i]:=ACK[i-1]+$(w_{i-1}-w_{i})$; \\
\> \} \\
\> else \{ \\
\> \> ACK[i]=ACK[i-1]; \\
\> \> if ($w_{i}<w_{i-1}$)  \\
\> \> \> LARGEST:= PACKET[i]:= LARGEST+$(w_{i}-w_{i})$;\\
\> \} \\
\> \} \\
for ($i=1$ to $n$)\{\\
\>  if ($w_{i}=w_{i-1}$)\\
\> \> let $PACKET[i]$ be the smallest positive integer \\
\> \> not present among values $PACKET[j]$, $1\leq j<i$;\\
\> \} \\
if (vector PACKET is a permutation of $\{1,\ldots, n\}$) \\
\> return PACKET; \\
else \\
\> return NO PERMUTATION EXISTS; \\
 } }
 \caption{Algorithm for reconstructing permutations from buffer sizes}\label{algrec}
\end{figure}

We prove the correctness of algorithm RECONSTRUCT in a couple of
intermediate steps. The first two apply to a general sequence $A$
(rather than one with $SUS(A)\leq 3$).

\begin{lemma}\label{aux}

Suppose there exists a permutation $\pi$ with ${\mathcal M}(\pi)=w$.
Then the following are true at any stage $i\geq 1$:
\begin{enumerate}
\item For any $j\geq 1$, the last element added to list $L_{j}$ is the maximum
element in lists $L_{k}$, $k\geq j$. In particular the largest
element of $L_{1}$ is the maximum element seen so far.
\item If element $x$ is the largest element seen up to stage $i$ then
$x=ACK_{i}+{\mathcal M}_{i}-1$.
\end{enumerate}
\end{lemma}

\beginproof
Let $i=1$. Statement 1. is clearly true. For the second statement,
note that $ACK_{1}=2$ and ${\mathcal M}_{1}=0$ if $x=1$ (in-order
packet) otherwise $ACK_{1}=1$, ${\mathcal M}_{1}=x$.

Consider now the case $i>1$. By the induction statement, the largest
element seen so far (call it $y$) is the last element of $L_{1}$ and
$y=ACK_{i-1}+{\mathcal M}_{i-1}$.

{\bf Case 1: $x$ is added to $L_{1}$}. By the definition $x>y$ so
$x$ is the largest element seen so far. Moreover, since $x$ is an
out-of-order element we have $ACK_{i}=ACK_{i-1}$ and ${\mathcal
M}_{i}={\mathcal M}_{i-1}+x-y$.

{\bf Case 2: $x$ is added to some other list $L_{j}$}. If $x$ is the
first element of the new list then statement 1 follows immediately.
Otherwise let $z$ be the largest element of list $L_{j}$ {\em
before} adding $x$. Applying the induction hypothesis it follows
that $z$ is the largest element in lists $L_{k}$, $k\geq j$. But
$z<x$ (since we add $x$ to list $L_{j}$). Thus $x$ becomes the new
largest element of lists $L_{k}$, $k\geq j$.

As for the second statement, from the algorithm it follows that
$x<y$ so $y$ is still the largest element seen so far. If the buffer
size does not modify then the desired relation follows from
$y=ACK_{i-1}+{\mathcal M}_{i-1}$ (which holds by induction) and
relations $ACK_{i}=ACK_{i-1}$ and ${\mathcal M}_{i}={\mathcal
M}_{i-1}$. Otherwise the buffer shrinks with size
$ACK_{i}-ACK_{i-1}$, so ${\mathcal M}_{i-1}-{\mathcal
M}_{i}=ACK_{i}-ACK_{i-1}$. We infer the fact that

\begin{eqnarray*}
y & = & ACK_{i-1}+{\mathcal M}_{i-1}-1=ACK_{i}-(ACK_{i}-ACK_{i-1})+{\mathcal M}_{i-1}-1=\\
& = & ACK_{i}+({\mathcal M}_{i}-{\mathcal M}_{i-1})+{\mathcal
M}_{i-1}-1=ACK_{i}+{\mathcal M}_{i}-1.
\end{eqnarray*}
\endproof

\begin{corollary}
Algorithm SUSGreedy correctly computes $u=LDS(A)$ (which is equal
\cite{levcopoulos-petersson-sus} to $SUS(A)$).
\end{corollary}
\beginproof
Let $B=a_{i_{1}}>a_{i_{2}}>\ldots > a_{i_{LDS(A)}}$ be a decreasing
subsequence of $W$ of maximum length, and let $L_{1},L_{2},\ldots,
L_{j}$ be the lists created by the algorithm on input sequence $A$.
Each list $L_{k}$ is increasing, so it contains at most one element
from $B$. Therefore $u\geq LDS(A)$. On the other hand, each element
$a_{m}$ set by the algorithm to a list $L_{k}$, $k\geq 2$ is smaller
than some element $a_{n}$, $n<m$, set by the algorithm to line $k-1$
(otherwise $a_{m}$ would be set to a list $L_{j}$, $j<k$). Applying
this observation starting with the last element of list $L_{u}$ we
create a decreasing sequence of length $u$. It follows that $u\leq
LDS(A)$, thus $u=LDS(A)$.
\endproof

From now on we assume that there exists a permutation $A$ with
$SUS(A)\leq 3$ such that ${\mathcal M}(A)=w$. We will run the
algorithm SUSgreedy along algorithm RECONSTRUCT. First we give a
simple corollary of Lemma~\ref{aux}:

\begin{corollary}\label{cor}
Suppose that $w_{i}> w_{i-1}$. Let $y$ be the largest ID of a packet
received in stages 1 to $i-1$ and $x$ be the ID of the new packet.
Then
\[
x=y+(w_{i}-w_{i-1})
\]

and $x$ is added by SUSgreedy to list $L_{1}$.
\end{corollary}

Next we deal with another possible case, the one when the buffer
size shrinks:

\begin{lemma}\label{foo}
 \begin{itemize}
 \item[(a).]
 Let packet ID $x$ be added at stage $i$, and assume that
 $w_{i}<w_{i-1}$
 Then $x=ACK_{i-1}$ and all packets with indices at most
 $ACK_{i-1}+(w_{i-1}-w_{i}-1)$ have been received in the first $i$
 stages.
 \item[(b).] Suppose packet ID $x$ is added by algorithm SUSgreedy to list
 $L_{3}$. Then
 packet $x$ falls into case (a) of this lemma.
 \end{itemize}
\end{lemma}
\beginproof
\begin{itemize}
\item[(a).] The fact that $x=ACK_{i-1}$ follows from the definition
of parameter $ACK$ and the fact that the buffer shrinks. The second
relation follows from the fact that the buffer shrinks by exactly
$w_{i-1}-w_{i}$. 
\item[(b).]
 Since $x$ goes in list $L_{3}$, at the time when added $x$ is
smaller than the last element in lists $L_{1}$ and $L_{2}$. If $x$
were larger than $ACK_{i-1}$ then the packet with index $ACK_{i-1}$
(which arrives sometimes after $x$ does) could not be placed in
lists $L_{1}$, $L_{2}$ or $L_{3}$, making the sequence $A$ require
$SUS(A)\geq 4$, a contradiction.

The other two relations follow from the definition of parameter
$ACK_{i}$.
\end{itemize}
\endproof

Finally, the correctness of the algorithm RECONSTRUCT (and the proof
of Theorem~\ref{main}) follows easily: the correctness of the first
for loop in algorithm RECONSTRUCT follows from Corollary~\ref{cor}
and Proposition~\ref{foo}. Moreover, if a packet ID $x$ is set at
stage $i$ in the second for loop then it must correspond to adding
$x$ to list $L_{2}$. Since list $L_{2}$ is sorted, $x$ is the
smallest element that has not been set up to this stage.

Assuming that permutation $A$ in the preimage of $w$ exists then
algorithm RECONSTRUCT is going to output exactly $A$. Since $A$ was
chosen in an arbitrary manner, the uniqueness of $A$ follows.

\endproof

\section{Application to RESTORED}

The result we just proved allows the reinterpretation of results in
\cite{restored,restored-metrics}. In  that paper it was shown
experimentally that \restored is able to recover several measures of
quality of service, among them the following metric
\cite{ietf-reorder-density}. For simplicity our version of the
metric is adapted to the case of permutations (i.e. sequences with
no repeats or packet losses):

\begin{definition}
{\bf Reorder Density (RD)}.

Consider an implementation-dependent parameter $DT$ that is a
positive integer or $\infty$. Given a permutation $\pi$ we define
the reorder density of $\pi$ as the distribution of {\em
displacements} $\pi[i]-i$, restricted to those displacements in the
range $[-DT,DT]$.
\end{definition}

We also need the following definition from \cite{restored-metrics}:

\begin{definition}\label{cons}
 A metric $M$ is {\em consistent with respect to $\equiv_{FB}$} if for any
two ID sequences $A$ and $B$,
\[
 A\equiv_{FB} B\implies M(A)=M(B).
\]
\end{definition}

In other words, a consistent measure $M$ takes equal values on
equivalent sequences.

\begin{example}
By equation~(\ref{maineq}), every measure defined in terms of the
time series of parameter RcwWindow (e.g. the average value of this
parameter) is consistent with respect to $\equiv_{FB}$.
\end{example}

In particular, since \restored\  (in the form used in
\cite{restored,restored-metrics}) guarantees that, on sequence $A$
it will reconstruct a sequence $R(A)$ such that $R(A)\equiv_{FB} A$,
it is not really that surprising that \restored\  should be able to
capture any metric consistent with respect to $\equiv_{FB}$. The
reason that the experimental results from \cite{restored} were
somewhat surprising is that RD is an example of an {\em inconsistent
measure} according to the terminology of Definition~\ref{cons}.

\begin{observation}
If $A=(4\mbox{ } 3\mbox{ } 2\mbox{ } 1)$ and $B=(4\mbox{ } 2\mbox{ }
3\mbox{ } 1)$ then the distributions of
displacements are $D(A)=\left(\begin{tabular}{cccc} -3 & -1 & 1 & 3 \\
1/4 &
1/4 & 1/4 & 1/4 \\
\end{tabular}\right)$ and $D(B)=\left(\begin{tabular}{ccc} -3 & 0 & 3 \\
1/4 &
1/2 & 1/4 \\
\end{tabular}\right)$, respectively. It is easy to see that, no
matter how we set the parameter $DT$ to either a positive integer or
$\infty$, the truncated versions of distributions $D(A), D(B)$ are
going to be different. Thus $A\equiv_{FB}(B)$ but $D(A)\neq D(B)$,
which means that measure RD is inconsistent independently of the
value of threshold parameter DT.
\end{observation}

However, Theorem~\ref{main} forces us to reevaluate this statement:
since the vast majority of traces used in \cite{restored} had
$SUS\leq 3$ the measure is "consistent in practice" (at least on
this dataset). Theorem~\ref{main} also exposes a weakness of the
encoding used in \cite{restored}: on "real-life" traces the extra
potential compression given by the many-to-one nature of map $FB$ is
not present.

\section{Acknowledgments}

I thank Anders Hansson for useful discussions.

\bibliographystyle{abbrv}

\bibliography{/home/gistrate/bib/bibtheory}

\end{document}